\documentclass[osajnl,amssymb,preprint]{revtex4}
\usepackage{psfig,graphicx}
\begin{document}
\title{Three-mode entanglement by interlinked nonlinear
interactions in optical $\chi^{(2)}$ media}
\author{Alessandro Ferraro, Matteo G. A. Paris}
\affiliation{Dipartimento di Fisica and Unit\`a INFM,
Universit\`a di Milano, Italy}
\author{Maria Bondani}
\affiliation{INFM, Unit\`a di Como, Italy}
\author{Alessia Allevi, Emiliano Puddu, Alessandra Andreoni}
\affiliation{INFM, Unit\`a di Como, Italy} \affiliation{
Dipartimento di Fisica e  Matematica, Universit\`a dell'Insubria,
Como, Italy}
\begin{abstract}
We address the generation of fully inseparable three-mode entangled states of
radiation by interlinked nonlinear interactions in $\chi^{(2)}$ media.  We show
how three-mode entanglement can be used to realize symmetric and asymmetric
telecloning machines, which achieve optimal fidelity for coherent states.  An
experimental implementation involving a single nonlinear crystal where the two
interactions take place simultaneously is suggested.  Preliminary experimental
results showing the feasibility and the effectiveness of the interaction
scheme with seeded crystal are also presented.
\end{abstract}
\maketitle
\section{Introduction}\label{s:intro}
The successful demonstration of continuous variable (CV) quantum
teleportation \cite{furu,bow,zha} and dense coding \cite{peng}
opened new perspectives to quantum information technology based on
Gaussian states of light. Besides having been recognized as the
essential resource for teleportation \cite{furu} and dense coding
\cite{ban}, the entanglement between two modes of light has been
proved as a valuable resource also for cryptography \cite{ralph,silb},
improvement of optical resolution \cite{fabre}, spectroscopy
\cite{spectr}, interferometry \cite{entdec}, state engineering
\cite{engstat}, and tomography of states and operations
\cite{tomoch,entame}.
\par
These achievements stimulated a novel interest in the generation
and application of multipartite entanglement
\cite{jing,zhang,aoki,pvl}, which has already received attention
in the domain of discrete variables. Multipartite CV entanglement
has been proposed to realize cloning at distance (telecloning)
\cite{telecl0,telecl1}, and to improve discrimination of quantum
operations \cite{gargnano}. The separability properties of CV
tripartite Gaussian states have been analyzed in \cite{geza},
where they have been classified into five different classes
according to positivity of the three partial transposes that can
be constructed. Moreover, it has been pointed out that genuine
applications of three-mode entanglement requires fully inseparable
tripartite entangled states \cite{vlb2000}, {\em i.e.} states that
are inseparable with respect to any grouping of the modes.
\par
Experimental schemes to generate multimode entangled states have
been already suggested and demonstrated. The first example,
although no specific analysis was made on the entanglement
properties (besides verification of teleportation), is provided by
the original teleportation experiments of Ref. \cite{furu} where
one party of a twin-beam (TWB) was mixed with a coherent state. A
similar scheme, where one party of a TWB is mixed with
the vacuum \cite{jing} has been demonstrated, and applied to
controlled dense coding. More recently, a fully inseparable
three-mode entangled state has been generated and verified
\cite{aoki} by mixing three independent squeezed vacuum states in
a network of beam splitters. In addition, a four-mode entangled
state to realize entanglement swapping with pulsed beams have been
generated \cite{leuch}.
\par
All the above schemes are based on  parametric sources, either of
single-mode squeezing or of two-mode entanglement {\em i.e.} TWB,
with multipartite entanglement resulting from further interactions
in linear optical elements ({\em e.g.} beam splitters). In
this paper, we focus on a scheme involving a single nonlinear
crystal, in which the three-mode entangled state is produced by
two type I, non-collinearly phase-matched interlinked bilinear
interactions that simultaneously couple the three modes
\cite{manuscript}. A similar interaction scheme, though realized
in type II collinear phase-matching conditions, is described in
Ref.~\cite{andrews70}. Compared to this work, our choice of
non-collinear phase-matching provides remarkable flexibility to
our experimental setup, whereas the choice of type I interaction
prevents the generation of additional parties. Moreover, we avoid
the losses brought about by the mode-matching in multiple beam
splitters in that we achieve the three-partite entanglement as
soon as we find the configuration that fulfills the phase-matching
condition for both interactions.
\par
The paper is structured as follows. In Section \ref{s:3gen} we
describe the generation of three-mode entanglement in a single
nonlinear crystal where two interlinked bilinear interactions take
place simultaneously. We obtain the explicit form in the Fock
basis of the outgoing three-mode entangled state, and also address
the characterization of entanglement. In Sections \ref{s:3cl} and
\ref{s:alpha} we show how the three-mode entangled state obtained
in our scheme, either for initial vacuum state or by seeding the
crystal, can be used to build symmetric and asymmetric telecloning
machines that achieve optimal fidelity for coherent states. In
Section \ref{s:TWB} we show how three-mode entanglement may be
used for conditional generation of two-mode entanglement, in
particular of TWB state. The scheme is of course less efficient
than direct generation of TWB in a parametric amplifier, but it
may be of interest in applications where {\em entanglement
on-demand} is required. In Section \ref{s:exp}, we discuss the
experimental implementation of our generation scheme. We show the
feasibility of experiments in the case of interaction with seeded
crystal and report preliminary experimental results. Section
\ref{s:conc} closes the paper with some concluding remarks.
\section{Generation of three-mode entanglement}\label{s:3gen}
The interaction Hamiltonian we are going to consider is given by
\begin{equation}
H_{int} =\gamma_1 a_1^\dag a^{\dag}_3 + \gamma_2 a_2^{\dag} a_3 + h.c.
\label{intH}\;.
\end{equation}
$H_{int}$ describes two interlinked bilinear interactions taking
place among three modes of the radiation field. It can be realized
in $\chi^{(2)}$ media by a suitable configuration which will be
discussed in Section \ref{s:exp}. The effective coupling constants
$\gamma_j$, $j=1,2$, of the two parametric processes are
proportional to the nonlinear susceptibilities and the pump
intensities. The Hamiltonian in Eq. (\ref{intH}) has been
firstly studied in \cite{SmithersLu74}, though not for the
generation of entanglement. The Hamiltonian admits the following
constant of motion
\begin{equation}
\Delta(t) \equiv N_1(t) - N_2(t) - N_3(t) \equiv \Delta (0)
\label{cm}\;,
\end{equation}
where $N_j(t)=\langle a^\dag_j(t) a(t)\rangle$ represent the
average number of photons in the $j$-th mode. If we take the
vacuum $|{\bf 0\rangle}\equiv |0\rangle_1 \otimes |0\rangle_2
\otimes |0\rangle_3$ as the initial state we have $\Delta=0$ {\em
i.e.} $N_1(t)=N_2(t)+N_3(t)$ $\forall t$. The expressions for
$N_j(t)$ can be obtained by the Heisenberg evolution of the field
operators, which read as follows
\begin{eqnarray}
a_1^\dag (t) &=& f_1 a_1^\dag (0) + f_2 a_2  (0) + f_3 a_3 (0) \nonumber \\
a_2 (t) &=& g_1 a_1^\dag (0) + g_2 a_2 (0) + g_3 a_3 (0) \nonumber \\
a_3 (t) &=& h_1 a_1^\dag (0) + h_2 a_2 (0) + h_3 a_3 (0)
\label{adit}\;.
\end{eqnarray}
The explicit expressions of the coefficients $f_j$, $g_j$ and $h_j$,
$j=1,2,3$, are obtained in appendix
\ref{a:hei}; we omit the time dependence for brevity.
By introducing $\Omega = \sqrt{|\gamma_2|^2 -|\gamma_1|^2}$ we have
\begin{eqnarray}
N_1 &=& N_2+N_3 \;, \nonumber \\
N_2 &=& \frac{|\gamma_1|^2 |\gamma_2|^2}{\Omega ^4}
\left[\cos{\Omega t}-1 \right]^2 \;, \nonumber \\
N_3 &=& \frac{|\gamma_1|^2}{\Omega ^2} \sin^2(\Omega t) \;.
\label{Ndit}
\end{eqnarray}
The evolved state reads as follows \cite{nic}
\begin{widetext}
\begin{equation}
|{\bf T}_0\rangle =U_t |{\bf 0}\rangle = \frac{1}{\sqrt{1+N_1}} \sum_{pq}
\left(\frac{N_2}{1+N_1}\right)^{p/2}
\left(\frac{N_3}{1+N_1}\right)^{q/2}
\sqrt{\frac{(p+q)!}{p! q!}}\: |p+q,p,q\rangle
\label{state}\;,
\end{equation}
\end{widetext}
where $U_t=\exp\left(-iH_{int}t\right)$ is the evolution operator, and we have
already used the conservation law.
The state in Eq. (\ref{state}) is Gaussian, as it can be easily demonstrated
by evaluating the characteristic function
\begin{eqnarray}
\chi(\lambda_1,\lambda_2,\lambda_3) &=& \hbox{Tr}\left[|{\bf T}_0\rangle\langle
{\bf T}_0|\: D_1(\lambda_1)\otimes D_2(\lambda_2)\otimes D_3(\lambda_3) \right]
\nonumber \\ &=& \langle {\bf 0}| U_t^\dag D_1(\lambda_1)\otimes D_2(\lambda_2)\otimes
D_3(\lambda_3) U_t
|{\bf 0}\rangle \nonumber \\
&=& \exp \left[-\frac12\left(|\lambda_1^\prime|^2+|\lambda_2^\prime|^2+
|\lambda_3^\prime|^2\right)\right]
\label{chfun}\;,
\end{eqnarray}
where $\lambda_j$ are complex numbers,
$D_j(\lambda_j)=\exp(\lambda_j a^\dag_j - \bar\lambda_j a_j)$ is a
displacement operator for the $j$-th mode, and the primed
quantities are obtained by using the Heisenberg evolution of the
modes in Eq.s~(\ref{adit}). In formulas
\begin{eqnarray}
\lambda_1^\prime &=& f_1 \lambda_1 - g_1 \overline{\lambda_2}
- h_1 \overline{\lambda_3} \nonumber \\
\lambda_2^\prime &=& - \overline{f_2} \overline{\lambda_1}
+ g_2 \lambda_2 + \overline{h_2} \lambda_3 \nonumber \\
\lambda_3^\prime &=& - \overline{f_3} \overline{\lambda_1}
+ \overline{g_3} \lambda_2 +h_3 \lambda_3
\label{lambdaprime}\;.
\end{eqnarray}
Following Ref. \cite{geza}, the characteristic function can be
rewritten as
\begin{eqnarray}
\chi(\lambda_1,\lambda_2,\lambda_3) = \exp \left[-\frac14
{\bf x}^T {\bf C} \: {\bf x} \right]
\label{introcov}\;,
\end{eqnarray}
where ${\bf x}^T=(x_1,x_2,x_3,p_1,p_2,p_3)$, $(\cdots)^T$
denotes transposition, $\lambda_j= 2^{-1/2} (p_j- i x_j)$,
$j=1,2,3$, and ${\bf C}$ denotes the covariance matrix of the
Gaussian state, whose explicit expression can be easily
reconstructed from Eq.s (\ref{lambdaprime}). The covariance matrix
determines the entanglement properties of $|{\bf T}_0\rangle$. In
fact, since $|{\bf T}_0\rangle$ is Gaussian the positivity of the
partial transpose is a necessary and sufficient condition for
separability \cite{geza}, which, in turn, is determined by the
positivity of the matrices $\Gamma_j=\Lambda_j {\bf C} \Lambda_j -
i {\bf J}$ where $\Lambda_1=\hbox{Diag}(1,1,1,-1,1,1)$,
$\Lambda_2=\hbox{Diag}(1,1,1,1,-1,1)$,
$\Lambda_3=\hbox{Diag}(1,1,1,1,1,-1)$ and ${\bf J}$ is the
symplectic block matrix
$$ \left(\begin{array}{cc}0 & - {\bf I} \\ {\bf I}
& 0\end{array}\right)\:, $$ ${\bf I}$ being the $3\times3$
identity matrix. A numerical evaluation of the eigenvalues of
$\Gamma_j$ shows that they are nonpositive matrices $\forall j$.
Correspondingly, the state in Eq.~(\ref{state}) is fully
inseparable {\em i.e.} not separable for any grouping of the
modes. Notice that the success of a true tripartite quantum
protocol, as the telecloning scheme described in the following
sections, is a sufficient criterion for the full inseparability of
the state $|{\bf T}_{0}\rangle$ \cite{vlb2000}.
\section{Telecloning of coherent states}\label{s:3cl}
\par\noindent
Here we show how the three-mode entangled state described in the
previous section can be used to achieve telecloning \cite{telecl0}
of coherent states \cite{telecl1}, that is to produce two clones
{\em at distance} of a given input radiation mode prepared in a
coherent state. Depending on the values of the coupling constants
of the Hamiltonian (\ref{intH}) the two clones can either be equal
one to each other or be different. In other words, the scheme is
suitable to realize both symmetric and asymmetric cloning machines
\cite{simasim}.  This option can eventually be useful to fit the
purpose of the clones production in order to distribute the
quantum information contained in the input state
\cite{clonfromeas,optclon1,optclon2,optclon3}.  Our scheme, which is analogous to
that of Ref. \cite{telecl1} in the absence of an amplification
process for the signal, is applied to the telecloning of coherent
states, whereas the state we use to support the teleportation is
the three-mode entangled state of Eq. (\ref{state}).  For the
symmetric case we obtain an optimal cloning machine, achieving the
maximum value of fidelity allowed in a continuous variable cloning
process ($F=2/3$) \cite{optclon1,optclon2,optclon3}. In the case of the asymmetric
cloning a range of coupling parameters can be found that allows
the fidelity of one clone to be greater than $2/3$, maintaining
the fidelity of the other greater than $1/2$, i.e. the maximum
value reachable in a classical communication scheme.
\begin{figure}[h!]
\includegraphics[width=0.35\textwidth]{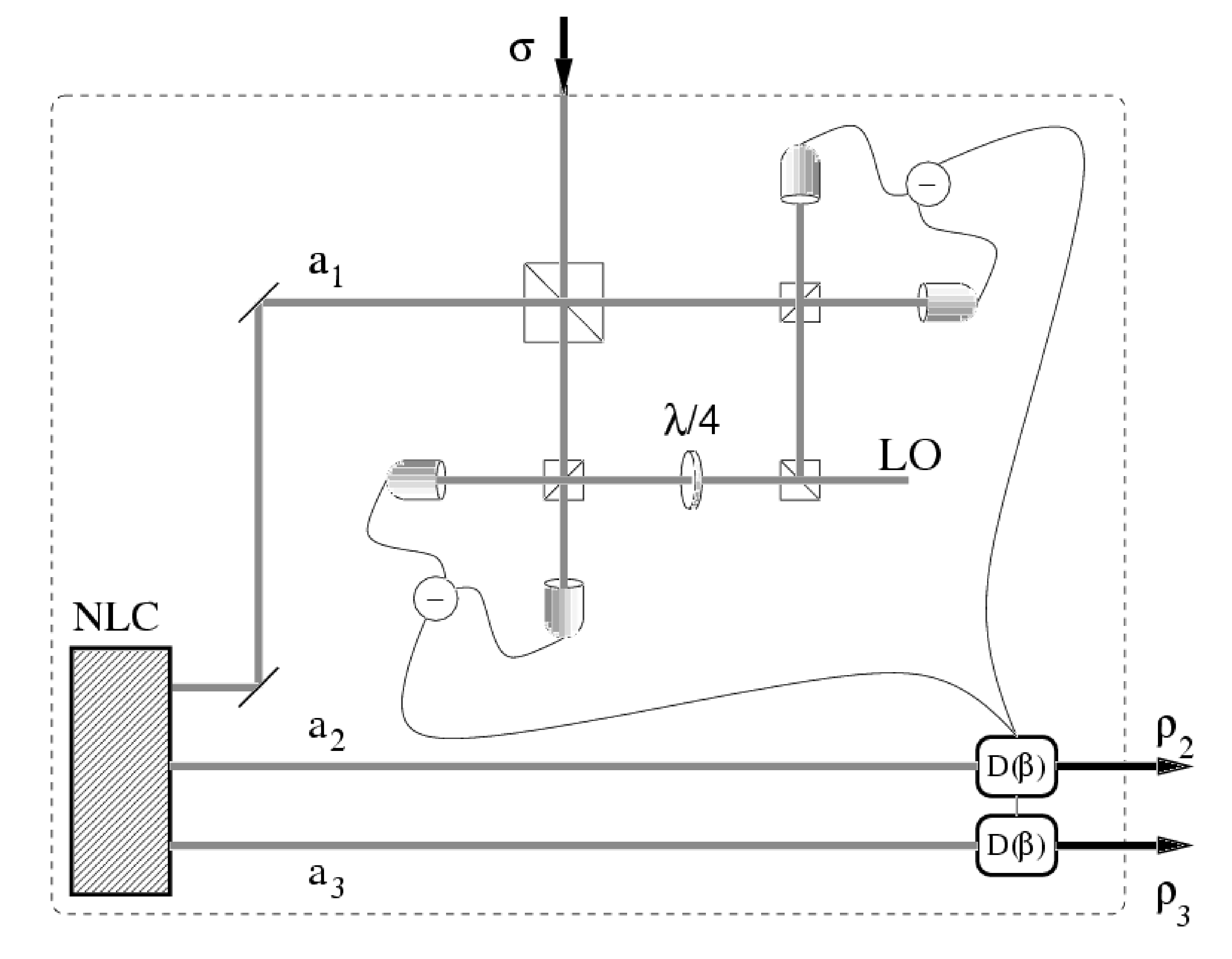}
\caption{\small Schematic diagram of the telecloning scheme. After the
preparation of the state $|{\bf T}_0\rangle$ by bilinear
interactions in a nonlinear crystal (NLC), a conditional
measurement is made on the mode $a_1$, which corresponds to the
joint measurement of the sum- and difference-quadratures on two
modes: mode $a_1$ itself and another reference mode $b$, which is
excited in a coherent state $\sigma$, to be teleported and cloned.
The result of the measurement is classically sent to the parties
who want to prepare approximate clones, where suitable
displacement operations (see text) are performed.
\label{f:telfig}}
\end{figure}
\par
A schematic diagram of our scheme is depicted in Fig. \ref{f:telfig}.
After the preparation of the state $|{\bf T}_0\rangle$ a conditional
measurement is made on the mode $a_1$, which corresponds to the joint
measurement of the sum- and difference-quadratures on two modes: mode $a_1$
and another reference mode $b$, whose state is to be teleported and cloned.
The measurement can be described as the following $\sigma$-dependent
POVM acting on the mode $a_1$
\begin{equation}
\Pi (\beta) = \frac{1}{\pi} D(\beta)\: \sigma^T D^\dag (\beta)
\label{povm}\;,
\end{equation}
where $D(\beta)$, $\beta \in {\mathbb C}$, is the displacement operator, and $\sigma$
is the preparation of $b$, {\em i.e.} the state to be teleported and cloned.
\par\noindent
The probability distribution of the outcomes is given by
\begin{widetext}
\begin{eqnarray}
P (\beta) &=& \hbox{Tr}_{123} \left[|{\bf T}_0\rangle\langle{\bf T}_0|\: 
\Pi(\beta) \otimes
\hbox{\bf I}_2 \otimes \hbox{\bf I}_3\right] \nonumber \\ &=& 
\frac{1}{\pi(1+N_1)} \sum_{pq}
\frac{N_2^p N_3^q}{(1+N_1)^{p+q}} \frac{(p+q)!}{p!\: q!} \:
\langle p+q|D(\beta) \sigma^T D^\dag (\beta) |p+q\rangle
\label{Palfa}\;.
\end{eqnarray}
The conditional state of the mode $a_2$ and $a_3$ after the outcome $\beta$
is given by
\begin{eqnarray}
\varrho_\beta &=& \frac{1}{P(\beta)}\: \hbox{Tr}_{1} \left[|{\bf T}_0
\rangle\langle{\bf T}_0|\:
\Pi(\beta) \otimes \hbox{\bf I}_2 \otimes \hbox{\bf I}_3\right] 
\nonumber \\ &=& \frac{1}{P(\beta)}\frac{1}{\pi(1+N_1)} \sum_{pqkl}
\frac{N_2^{(p+k)/2} N_3^{(q+l)/2}}{(1+N_1)^{(p+q+k+l)/2}}\:
\sqrt{\frac{(p+q)!\:(k+l)!}{p!\: q!\:k!\:l!}} \nonumber \\ &\times&
\langle k+l|D(\beta) \sigma^T D^\dag (\beta) |p+q\rangle \: |p,q\rangle\langle
k,l|
\label{Rhoalfa}\;.
\end{eqnarray}
\end{widetext}
After the measurement the conditional state may be transformed by a
further unitary operation, depending on the outcome of the
measurement. In our case, this is a two-mode product displacement
$U_\beta = D^T(\beta)\otimes D^T(\beta)$ where the amplitude $\beta$ is
equal to the results of the measurement. This is a local transformation
which generalizes to two modes the procedure already used in the
original CV $1\rightarrow 1$ teleportation protocol. The overall state
of the two modes is obtained by averaging over the possible outcomes
$$
\varrho_{23}=\int_{\mathbb C} d^2\beta\: P (\beta) \:
\tau_\beta\:.$$
where $\tau_\beta=U_\beta\: \varrho_\beta\:
U_\beta^\dag$.
\par
\begin{figure}[h]
\includegraphics[width=0.35\textwidth]{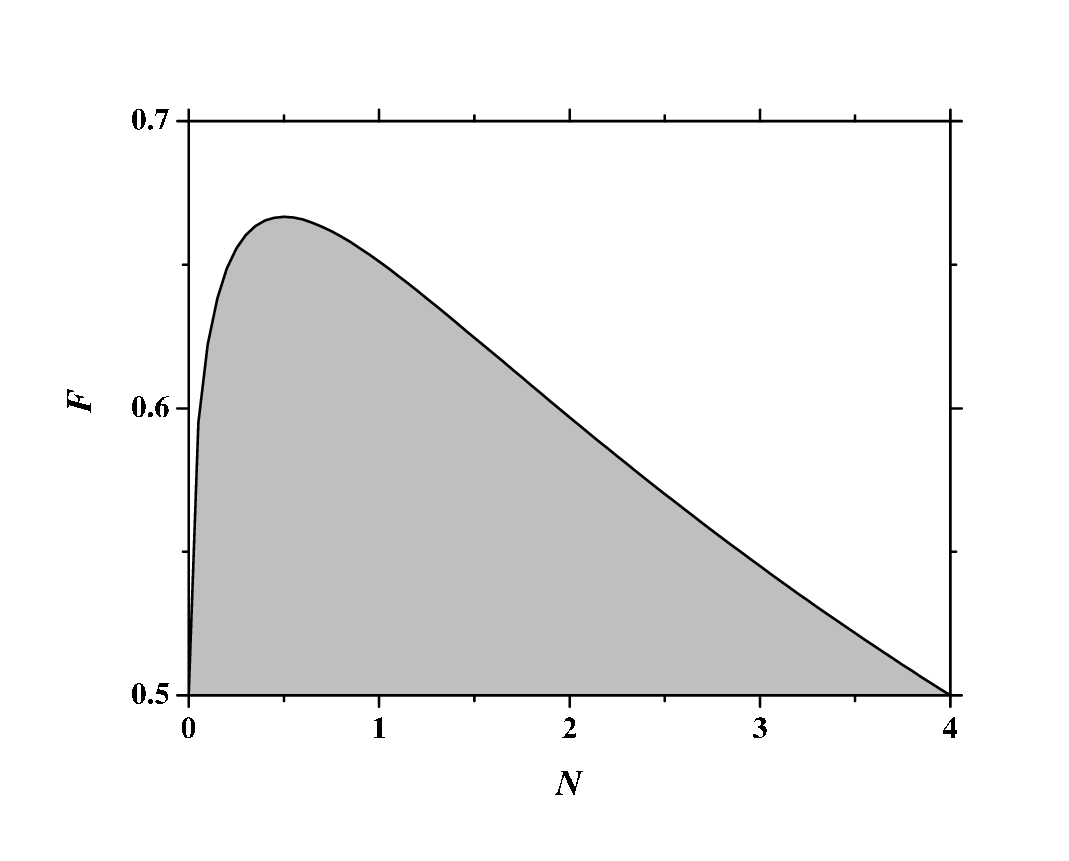}
\caption{\label{f:fid} Fidelity of symmetric clones versus the
average (equal) photon number $N$ of modes $a_2$ and $a_3$.}
\end{figure}

\noindent
If $b$ is excited in a coherent state $\sigma=|z\rangle\langle z|$ the
probability of the outcomes is given by
\begin{equation}
P_z(\beta) = \frac{1}{\pi(1+N_1)} \:
\exp \left\{-\frac{|\beta+\overline{z}|^2}{1+N_1}\right\}
\label{PalfaCoh}\;.
\end{equation}
Moreover, since the POVM is pure also the conditional state is pure.
We have $\varrho_\beta= |\psi_\beta\rangle\rangle\langle\langle
\psi_\beta|$ with
\begin{equation}
|\psi_\beta\rangle\rangle = |\delta_{2\beta}\rangle_2 \otimes
|\delta_{3\beta}\rangle_3 \label{PsiBeta}\;,
\end{equation}
{\em i.e.} the product of two independent coherent states. The
amplitudes are given by
$$
\delta_{2\beta}=(z +\overline{\beta})\kappa_2 \qquad \delta_{3\beta}= (z
+\overline{\beta})\kappa_3\:,$$
where the quantities $\kappa_j$, $j=2,3$ are
given by
\begin{equation}
\kappa_j = \sqrt{\frac{N_j}{1+N_1}}
\label{kappas}\;.
\end{equation}
Correspondingly, we have $
\tau_\beta=U_\beta\:|\psi_\beta\rangle\rangle\langle\langle
\psi_\beta| \: U_\beta^\dag $ with
\begin{equation}
U_\beta\:|\psi_\beta\rangle\rangle=
|z\kappa_2+\overline{\beta}(\kappa_2-1)\rangle \otimes
|z\kappa_3+\overline{\beta}(\kappa_3-1)\rangle \:.
\label{out33}\end{equation}
The partial traces
$\varrho_2=\hbox{Tr}_3[\varrho_{23}]$ and
$\varrho_3=\hbox{Tr}_2[\varrho_{23}]$ read as follows
\begin{eqnarray}
\varrho_j = \int_{\mathbb C} d^2\beta\: P_z (\beta) \:
|z\kappa_j+\overline{\beta}(\kappa_j-1)\rangle\langle z\kappa_j+\overline{\beta}(\kappa_j-1) |
\label{clones}\;.
\end{eqnarray}
We see from the teleported states in Eq.~(\ref{clones}) that it is
possible to engineer a symmetric cloning protocol if $N_2=N_3=N$,
otherwise we have an asymmetric cloning machine. Consider first
the symmetric case. According to Eq.s (\ref{Ndit}) the condition
$N_2=N_3=N$ holds when
\begin{equation}
\cos{\Omega t} = \frac{|\gamma_1|^2}{2|\gamma_2|^2-|\gamma_1|^2}
\label{condition}
\end{equation}
from which it follows that
\begin{equation}
N = \frac{4|\gamma_1|^2|\gamma_2|^2}{(2|\gamma_2|^2-|\gamma_1|^2)^2}
\label{Nequal}\;.
\end{equation}
Since
$|\langle\beta^{'}|\beta^{''}\rangle|^2=\exp\{-|\beta^{'}-\beta^{''}|^2\}$,
the fidelity of the clones is given by
\begin{widetext}
\begin{eqnarray}
F=\langle z|\varrho_j|z\rangle  &=& \int_{\mathbb C} \frac{d^2\beta}{\pi (2N+1)}\:
\exp\left\{-\frac{|z+\overline{\beta}|^2}{2N+1}\right\}\:
\exp\left\{-|z+\overline{\beta}|^2 (\kappa-1)^2\right\} 
\nonumber \\ &=& \frac{1}{2 + 3N - 2\sqrt{N(2N+1)}}
\label{fid}\;.
\end{eqnarray}
\end{widetext}
As we expect from a proper cloning machine, the fidelity is
independent of the amplitude of the initial signal and for $0<N<4$
it is larger than the classical limit $F=1/2$. Notice that the
transformation $U_\beta$ performed after the conditional
measurement, is the only one assuring that the output fidelity is
independent of the amplitude of the initial state. In Fig.
\ref{f:fid} the behavior of the fidelity versus the average photon
number $N$ is shown in the relevant regime. We can see that the
fidelity reaches its maximum $F=2/3$ for $N=1/2$ which means,
according to Eq. (\ref{Nequal}), that the physical system allows
an optimal cloning when its coupling constants are chosen so that
$|\gamma_1/\gamma_2|=\sqrt{6-\sqrt{32}}\simeq 0.586$ . Let us now
consider the asymmetric case. For $N_2 \neq N_3$ the fidelities 
$F_j = \langle z|\varrho_j|z\rangle$
of the two clones (\ref{clones}) are given by
\begin{eqnarray}
F_2 = 
\frac{1}{2 + N_3 +2N_2 - 2\sqrt{N_2(N_2+N_3+1)}}
\label{FidAsym1}\\
F_3 = \frac{1}{2 + N_2 +2N_3 - 2\sqrt{N_3(N_2+N_3+1)}}
\label{FidAsym2}\;.
\end{eqnarray}
A question arises whether it is possible to tune the coupling
constants so as to obtain a fidelity larger than the bound $F=2/3$
for one of the clones, say $\varrho_2$, while accepting a
decreased fidelity for the other clone. Indeed, for example, if we
impose $F_3=1/2$, {\em i.e.} the minimum value to assure the
genuine quantum nature of the telecloning protocol, then we should
choose $N_3=\frac14 N^2_2$. In this case the maximum value for
$F_2$ is given by $F_{2max}=4/5$, which occurs for $N_2=1$. More
generally, by substituting Eq. (\ref{FidAsym2}) in Eq.
(\ref{FidAsym1}) and maximizing $F_2$ with respect to $N_2$
keeping $F_3$ fixed, we obtain that for $N_2=1/F_3-1$ and
$N_3 = 1/4 (1/F_3-1)^{-1}$ we have
$$
F_2 = 4 \frac{(1-F_3)}{(4-3 F_3)}\:, $$ which shows that $F_2$ is
a decreasing function of $F_3$ and that $2/3 <F_2<4/5$ for $1/2
<F_3< 2/3 $. Notice that the sum of the two fidelities $F_2+F_3=1+
3/4 F_2 F_3$ is not constant, being maximum in the symmetric case
$F_2=F_3=2/3$. Notice also that the roles of $\varrho_2$ and
$\varrho_3$ are interchangeable in the considerations.
\section{Telecloning with seeded crystal}
\label{s:alpha} In order to confirm the feasibility of the
telecloning scheme presented in the previous section we now show
that the same protocol can be implemented also when the state that
supports teleportation is generated by Hamiltonian (\ref{intH})
starting from a coherent state in one of the modes, rather than
from the vacuum. This may be of interest from the experimental
point of view, since seeding a crystal with a coherent beam is a
useful technique to align the setup, and allows the verification
of the classical evolution of the interacting fields
[see Section~\ref{s:exp}].
\par
The analysis of the scheme is analogue to that of the previous
Section, however starting from the initial state
$|\alpha,0,0\rangle$ instead of the vacuum.
The explicit expression of the evolved state $|{\bf T}_\alpha\rangle$
is derived in appendix \ref{a:SchAlpha}.
Notice that the conservation law (\ref{cm}) implies that 
the populations for seeded crystal $N_{j\alpha}=\langle 
{\bf T}_\alpha |a^\dag_j a_j|{\bf T}_\alpha \rangle$
satisfies the relation
$N_{1\alpha}-N_{2\alpha}-N_{3\alpha}=|\alpha|^2$.
We refer the reader to appendix \ref{a:SchAlpha}
for the explicit expressions of $N_{j\alpha}$ and 
for their connections to the populations $N_j$ for 
vacuum input.  \par
A compact expression for the evolved state is the following
\begin{equation}
  \label{T_alpha_D}
|{\bf T}_\alpha\rangle = D_1(\alpha f_1(-t))\otimes D_2(-
\overline{\alpha f_2(-t)}) \otimes D_3(-
\overline{\alpha f_3(-t)}) |{\bf T}_0\rangle\:,
\end{equation}
where the $f_j(t)$, $j=1,2,3$ are given in appendix \ref{a:hei}.
Expression (\ref{T_alpha_D}) can be easily derived by using the
Heisenberg equation of motion for the field-mode $a_1(t)$ (see
Eq.s (\ref{adit})).
The telecloning process proceed as in the previous Section, 
with calculations performed using the shifted Fock basis
$|\psi_n\rangle_1 \equiv D_1(\alpha f_1)|n\rangle_1$, and
$|\psi_n\rangle_j \equiv D_j(-\overline{\alpha}
\overline{f}_j)|n\rangle_j$, $j=2,3$.
If the reference mode $b$ is excited in a pure coherent state
$\sigma=|z\rangle\langle z|$,
then, as in Section \ref{s:3cl}, the conditional state is pure
$\varrho_\beta= |\psi_\beta\rangle\rangle\langle\langle\psi_\beta|$
with
\begin{equation}
|\psi_\beta\rangle\rangle = |\zeta_{2\beta}\rangle_2 \otimes
|\zeta_{3\beta}\rangle_3 \label{PsiBetaAlfa}\;,
\end{equation}
{\em i.e.} the product of two independent coherent states. The
amplitudes are given by
$$
\zeta_{2\beta}=(z +\bar\beta-\overline{\alpha f_1})\kappa_2
-\overline{\alpha f_2} \qquad \zeta_{3\beta}=(z
+\bar\beta-\overline{\alpha f_1})\kappa_3 -\overline{\alpha
  f_3}\:,$$
where the quantities $\kappa_j$, $j=2,3$ are given by Eq.
(\ref{kappas}).  
The unitary transformation on $a_2$ and $a_3$ that 
completes the telecloning is now given by 
\begin{equation}
  \label{UdiBetaAlpha}
  U_\beta = D_2^\dag(\bar\beta-\kappa_2\overline{\alpha f_1}-
  \overline{\alpha f_2})\otimes D_3^\dag(\bar\beta-\kappa_3
  \overline{\alpha f_1}-\overline{\alpha f_3})\:.
\end{equation}
In fact, the output conditional state coincides with that of Eq.
(\ref{out33}), so that the partial traces are identical to those
given in Eq. (\ref{clones}). For $N_2=N_3=N$ we obtain symmetric
clones with the same fidelity as in Section \ref{s:3cl}. Moreover
conditions (\ref{condition}) and (\ref{Nequal}) still hold. Notice
that also the protocol for asymmetric cloning can be
straightforwardly extended to the present seeded scheme.
\section{Conditional generation of two-mode entanglement} \label{s:TWB}
Another application of the three-mode entangled state of Eq.
(\ref{state}) is the conditional generation of a two-mode entangled
state of radiation by on-off photodetection on one of the modes of
state $|{\bf T}_0\rangle $. Indeed, it is possible to produce a robust
two-mode entangled state that approaches a TWB for unit quantum
efficiency $\eta$ of the photodetector. In the following we evaluate
some properties of the conditional state when $\eta \neq 1$ in order
to quantify its closeness to an ideal TWB.  Notice that, due to the
well known properties of TWB, this scheme also provides a valid check
of the whole apparatus from an experimental viewpoint. Let us consider
the situation in which a mode of the state $|{\bf T}_0\rangle $, say
the third mode, is revealed by an on-off photodetector. The
probability operator measure (POVM) is two-valued $\{\Pi_0,\Pi_1\}$,
$\Pi_0+\Pi_1={\bf I}$, with the element associated to the "no photons"
result given by
\begin{equation}
\label{NoPh} \Pi_0= \hbox{\bf I}_1 \otimes \hbox{\bf I}_2
\otimes \sum_{n} (1-\eta)^n |n\rangle_3{}_3\langle n| \:.
\end{equation}
The probability of the outcome is given by
\begin{eqnarray}
  \label{PZero} P_0 &=& \hbox{Tr}_{123} \left[|{\bf T}_0\rangle\langle{\bf T}_0|\:
  \Pi_0\right] \nonumber \\ &=& \frac{1}{1+N_1}\sum_{m,n}
  \left(\frac{N_2}{1+N_1}\right)^n \left(\frac{N_3(1-\eta)}{1+N_1}\right)^m
  \frac{(n+m)!}{n!m!} \nonumber \\ &=& (1+\eta N_3)^{-1} \:,
\end{eqnarray}
while the conditional output state $\varrho_0 = \frac{1}{P_0}
  \hbox{Tr}_3\left[|{\bf T}_0\rangle\langle{\bf T}_0|\:
  \Pi_0\right]$
reads as follows
\begin{widetext}
\begin{eqnarray}
  \label{RoZero} \varrho_0 =  
  \frac{1+\eta N_3}{1+N_1} \sum_{m,n,n'}
  \left(\frac{N_2}{1+N_1}\right)^{\frac{n+n'}{2}}\left(\frac{N_3(1-\eta)}{1+N_1}\right)^m
  \frac{1}{m!}\sqrt{\frac{(n+m)!(n'+m)!}{n!n'!}} 
  |n+m,n\rangle\langle n'+m,n'|\:.
  \end{eqnarray}
Remind that $N_1=N_2+N_3$.  If $\eta=1$ this state reduces to
the following TWB
\begin{equation}
  \label{TwinBeam}
  |\psi_0\rangle=\sqrt{\frac{1+ N_3}{1+N_1}} \sum_{n}
  \left(\frac{N_2}{1+N_1}\right)^{\frac{n}{2}}|n,n\rangle\:.
\end{equation}
When the efficiency of the detector is not unitary a question
arises on how to quantify the closeness of $\varrho_0$ to the
ideal state $|\psi_0\rangle$. From an operational point of view,
we can evaluate the photon number correlation between the first
and second mode, which is defined as
\begin{equation}
  \label{PhCorrDef}
  \zeta_{12}=\frac{\langle(n_1-n_2)^2\rangle-(\langle n_1
  \rangle -\langle n_2\rangle)^2}{\langle n_1\rangle+\langle n_2\rangle}\:,
\end{equation}
and is zero in case of TWB.
After straightforward calculations we arrive at
\begin{eqnarray}
  \label{Zeta12}
  \zeta_{12}  =
  \frac{N_3(1-\eta)(1+N_3)}{(1+\eta N_3)[2N_2+N_3(1-\eta)]} \:,
\end{eqnarray}
which, for any given value of the quantum efficiency $\eta$, is a
decreasing function of $N_2$ and an increasing function of $N_3$.
A global quantity to characterize the state in Eq.~(\ref{RoZero})
is the fidelity with a reference TWB state. The natural choice for
the reference is the TWB $|\psi_0\rangle$, according to the
following argument. At first we calculate the fidelity between
state (\ref{RoZero}) and a generic TWB of parameter $\xi$
{\em i.e.} $|\xi\rangle=\sqrt{1-\xi^2}\sum\xi^n|n,n\rangle$, we
have
\begin{eqnarray}
  \label{FidelityTWBxi}
  F(\eta,\xi) &=& \langle\xi|\varrho_0|
  \xi\rangle \: = \: (1-\xi^2)\frac{1+\eta N_3}{1+N_1} \nonumber \\ &\times& 
  \sum_{m,n,n',p,q} \xi^{p+q}\left(\frac{N_2}{1+N_1}\right)^{\frac{n+n'}{2}}
  \left(\frac{N_3(1-\eta)}{1+N_1}\right)^m 
  \frac{1}{m!}\sqrt{\frac{(n+m)!(n'+m)!}{n!n'!}}
  \delta^p_n\delta^q_{n'}\delta^m_0 \nonumber
  \\ &=&\frac{1+\eta N_3}{1+N_1}\frac{1-\xi^2}{\left(1-\xi\sqrt{N_2/\left(1+N_1\right)}\right)^2} \:.
\end{eqnarray}
\end{widetext}
Then, we look for the parameter $\xi$ that maximizes the fidelity.
Expression (\ref{FidelityTWBxi}) shows that the value of $\xi$
maximizing the fidelity is, independently on $\eta$,
$\xi=\sqrt{N_2/(1+N_1)}$. By substituting in
Eq.~(\ref{FidelityTWBxi}) we arrive at
\begin{eqnarray}
  \label{FidelityTWB}
  F(\eta)=\frac{1+\eta N_3}{1+N_3} \:.
\end{eqnarray}
Therefore, the maximum fidelity is obtained for $\eta=1$ and the
correct reference state is the TWB $|\psi_0\rangle$. In
conclusion, the state generated through conditional on/off
photodetection on the third mode of $|{\bf T}_0\rangle$ is a
robust two-mode entangled state with a fidelity to a TWB given by
(\ref{FidelityTWB}). Notice that $\eta<F(\eta)<1$ for any choice
of $N_3$. The same analysis is valid for a conditional measurement
performed on mode $a_2$, in which case we obtain an entangled
state of modes $a_1$ and $a_3$ [in this case the role of $N_2$ and
$N_3$ should be exchanged in Eq.s (\ref{Zeta12}),
(\ref{FidelityTWBxi}), and (\ref{FidelityTWB})]. On the other
hand, we notice that a conditional photodetection on mode $a_1$
does not lead to an entangled state of modes $a_2$ and $a_3$.
\section{The optical scheme}\label{s:exp}
An experimental implementation of the scheme proposed in this
paper can be obtained by using a single nonlinear crystal in which
the two interactions described by Hamiltonian (\ref{intH}) take
place simultaneously. The interactions correspond to two
phase-matched second-order nonlinear processes in which five
fields interact and two of them do not evolve (parametric
approximation). Among the five fields $a_j$ involved in the
interactions, $a_4$ and $a_5$ will be the non-evolving
pump-fields. 
\begin{figure}[h]
\includegraphics[width=0.4\textwidth]{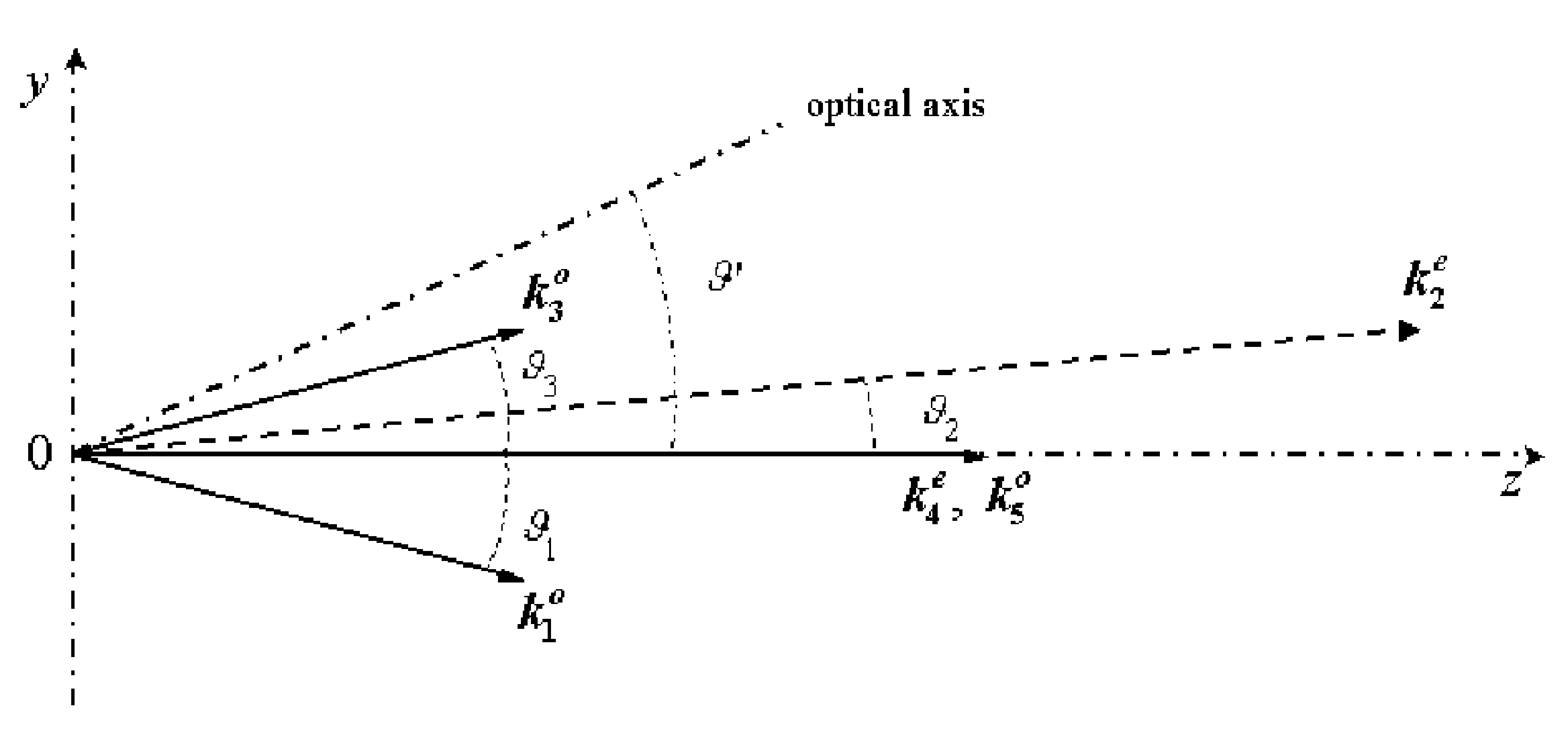}
\caption{\label{f:inter} Interaction scheme. The pump beams $a_4$
and $a_5$ are supposed to impinge on the crystal face along the
normal. The values of the crystal cut angle, $\vartheta'$, and of
the interaction angles $\vartheta_1$, $\vartheta_2$, and
$\vartheta_3$, are calculated to satisfy the phase-matching
conditions. The wavelengths of the interacting modes are
$\lambda(\omega_1) = \lambda(\omega_3) = 1064$ nm,
$\lambda(\omega_4) = \lambda(\omega_5) = 532$ nm and
$\lambda(\omega_2) = 355$ nm.}
\end{figure}
\par
\begin{figure}[h]
\includegraphics[width=0.4\textwidth]{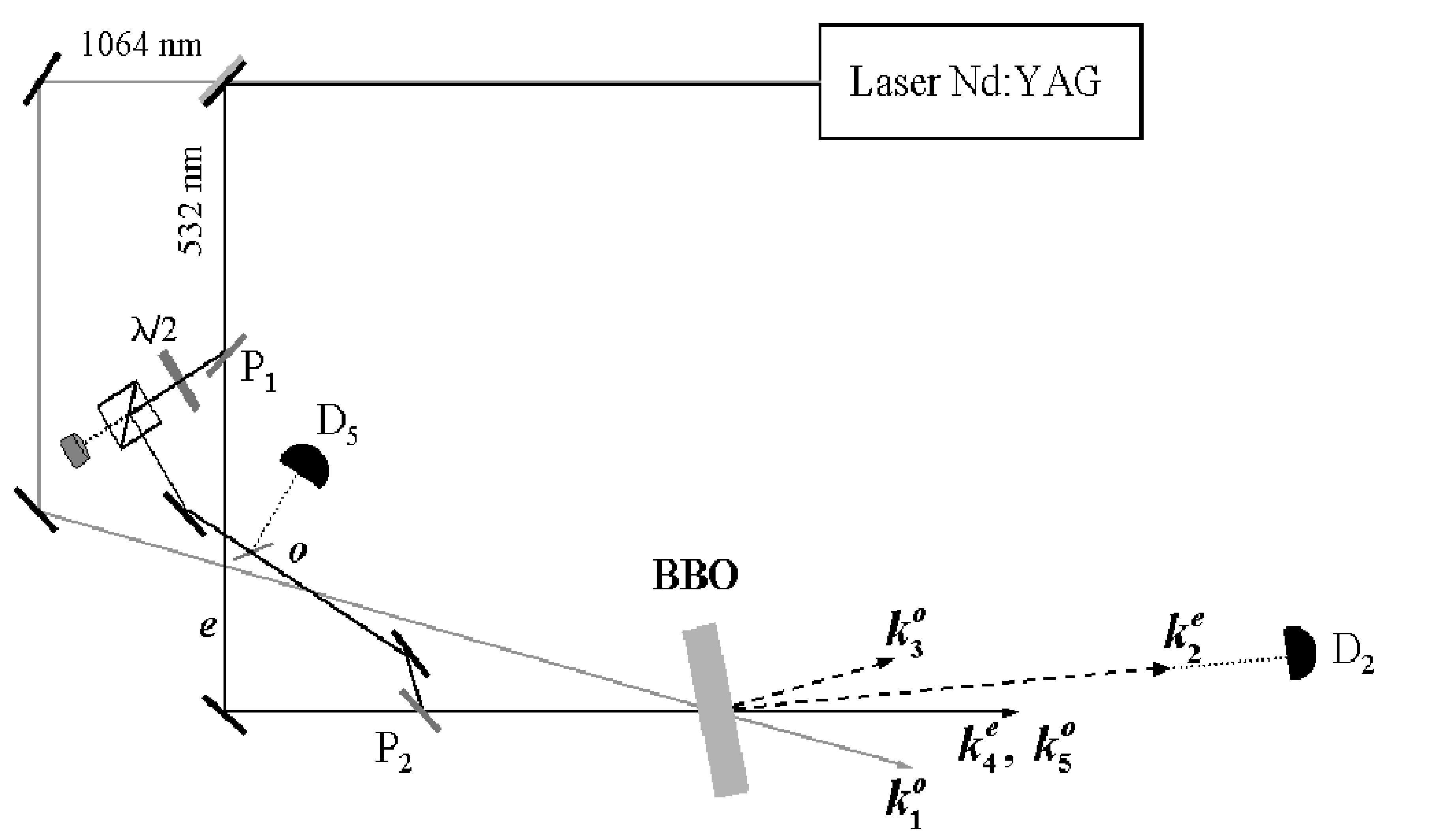}
\caption{\label{f:setup} Experimental setup. The fundamental and
second harmonic outputs of the Nd:YAG laser are sent to a harmonic
separator and then each beam is collimated to a diameter suitable
to illuminate the BBO crystal. The polarization of the second
harmonic beam emerging from the laser is elliptic, and the two
polarization components are separated by a thin-film plate
polarizer ($\mathrm{P}_1$). On the ordinarily polarized component
a $\lambda/2$ plate is inserted to modulate the intensity of beam
$a_5$, without affecting the intensity of the other pump, $a_4$.
The two beams then recombine at a second thin-film plate polarizer
($\mathrm{P}_2$) and are sent to the BBO.  $D_5$ and $D_2$ are
pyroelectric detectors.}
\end{figure}

The energy-matching and phase-matching conditions
required by the interactions can be written as $\omega_4 =
\omega_1 + \omega_3$, $\omega_2 = \omega_3 + \omega_5$,
$\mathbf{k}_4 = \mathbf{k}_1 + \mathbf{k}_3$ and $ \mathbf{k}_2 =
\mathbf{k}_3 + \mathbf{k}_5$, being $\mathbf{k}_j$ the
wave-vectors (in the medium) corresponding to $\omega_j$, which
make angles $\vartheta_j$ with the normal to the entrance face of
the crystal. It is possible to satisfy these phase-matching
conditions with a number of different choices of frequencies and
interaction angles depending on the choice of the nonlinear
medium. Here we propose an experimental setup based on a
$\beta-\rm{BaB}_2 \rm{O}_4$ crystal (BBO, cut angle 32 deg, cross
section $10\times 10$ mm$^2$ and 4 mm thickness, Fujian Castech
Crystals Inc., Fuzhou, China) as the nonlinear medium and the
harmonics of a Q-switched amplified Nd:YAG laser (7 ns pulse
duration, Quanta-Ray GCR-3-10, Spectra-Physics Inc., Mountain
View, CA) as the interacting fields. We choose a compact
interaction geometry in which two type I non-collinear
interactions with the two pump-beams superimposed in a single beam
with mixed polarization take place (see Fig. \ref{f:inter}). With
reference to Fig.~\ref{f:inter}, the wavelengths of the
interacting modes are $\lambda(\omega_1) = \lambda(\omega_3) =
1064$ nm, $\lambda(\omega_4) = \lambda(\omega_5) = 532$ nm and
$\lambda(\omega_2) = 355$ nm. The interaction angles, calculated
by supposing that the two pump beams propagate along the normal to
the crystal entrance face, result to be $\vartheta' = 37.74$ deg ,
$\vartheta_{1} = -\vartheta_{3} = 10.6$ deg and $\vartheta_{2} =
3.5$ deg, and since the crystal we used was cut at 32 deg, it had
to be rotated to allow phase matching. In order to demonstrate the
feasibility of the scheme in Fig.~\ref{f:inter}, we adopted the
experimental setup depicted in Fig.~\ref{f:setup}. The fundamental
and second harmonic outputs of the Nd:YAG laser were sent to a
harmonic separator and then each beam was collimated to a diameter
suitable to illuminate the BBO crystal. The polarization of the
second harmonic beam emerging from the laser is elliptic, and the
two polarization components were separated through a thin-film
plate polarizer ($\mathrm{P}_1$ in Fig.~\ref{f:setup}). On the
ordinarily polarized component a $\lambda/2$  plate was inserted
to modulate the intensity of beam $a_5$, without affecting the
intensity of the other pump, $a_4$. The two beams were then
recombined through a second thin-film plate polarizer
($\mathrm{P}_2$) and sent to the BBO. As a first verification of
the effectiveness of the interaction described by the Hamiltonian
(\ref{intH}), we implemented the seeded configuration discussed in
Section~\ref{s:alpha} by injecting the BBO with a portion of the
fundamental laser output (see Fig.~\ref{f:setup}) to realize the
initial condition for field $a_1$.
\par
\begin{figure}[h]
\includegraphics[width=0.4\textwidth]{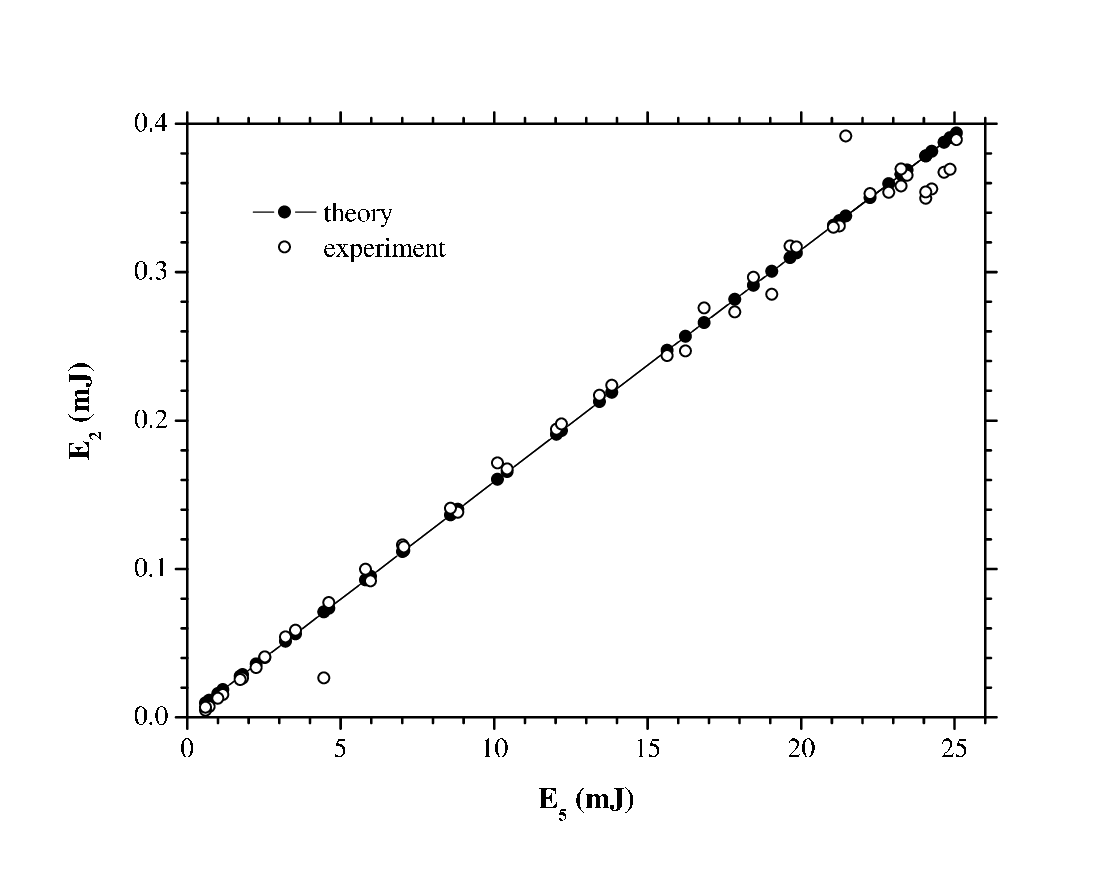}
\caption{\label{f:measure}
Comparison of the experimental results with the field evolution
calculated according to equation (\ref{eq:solclass}).
Open circles: measured values of the
energy of field $a_2$ as a function of the measured values of the
energy of pump-field $a_5$. Full circles: values of the energy of
field $a_2$ as calculated from the classical evolution of the
interacting fields as a function of the measured values of the
energy of pump-field $a_5$.}
\end{figure}

As a first quantitative check, we measured the energy,
$\mathrm{E}_2$, of the beam generated at $\omega_2$ as a function
of the energy, $\mathrm{E}_5$, of the ordinarily polarized
pump-beam for fixed values of the energies of the extraordinarily
polarized pump-beam, $\mathrm{E}_4$, and of the seed-beam,
$\mathrm{E}_1$. We preliminarily measured $\mathrm{E}_1$ by using
a pyroelectric detector (ED200, Gentec Electro-Optics Inc.,
Quebec, QC, Canada) which also allows checking the stability of
the source. By averaging over more than 100 pulses we found a
value of about 48 mJ per pulse, only $50\%$ of which is ordinarily
polarized, and thus suitable for the interaction. To measure
energy $\mathrm{E}_4$ we inserted another pyroelectric detector
(mod. ED500, Gentec) after $P_2$. By averaging again over more
than 100 pulses we found a value of about 158 mJ per pulse. To
obtain a reliable measurement of $\mathrm{E}_5$ we inserted, on
the path of beam $a_5$, a cube beam splitter and a calibrated
glass plate to extract a fraction of the beam. Energy
$\mathrm{E}_5$ was varied by rotating the $\lambda/2$ plate, and
its measurement was performed with the same detector ED500 as
before (see $\mathrm{D}_5$ in Fig.~\ref{f:setup})). To measure the
energy $\mathrm{E}_2$ of the output pulses we used another
pyroelectic detector (PE10, Ophir Optronics Ltd., Jerusalem,
Israel, see $\mathrm{D}_2$ in Fig.~\ref{f:setup})). The values of
$\mathrm{E}_5$ and $\mathrm{E}_2$ were measured simultaneously as
averages over the same 20 laser shots at each rotation of the
$\lambda/2$ plate. In Fig~\ref{f:measure} we show the measured
values of $\mathrm{E}_2$ (open circles), as a function of the
measured values of $\mathrm{E}_5$. We can compare the experimental
results with the field evolution calculated according to the
classical equations \cite{manuscript}
\begin{eqnarray} E_2 &=&
\frac{\omega_2}{\omega_1}\frac{c_1 E_4 \cdot c_2 E_5}{\left(c_2
E_5 - c_1 E_4\right)^2} \left[\cos\left(\sqrt{c_2 E_5 - c_1 E_4}\
z\right) - 1\right]^2 E_1\label{eq:solclass}
\end{eqnarray}
where $c_1 = 8.3\times 10^4 (\mathrm{Jm}^2)^{-1}$ and $c_2 =
2.6\times 10^5 (\mathrm{Jm}^2)^{-1}$ are the coupling constants
that apply to the present interactions. In Fig~\ref{f:measure} we
show the values (full circles) of $\mathrm{E}_2$ as calculated
according to Eq.~(\ref{eq:solclass}) for the experimental values
of $\mathrm{E}_5$, and for fixed values $\mathrm{E}_1 = 24$ mJ and
$\mathrm{E}_4 = 158$ mJ. The agreement between measured and
calculated values is excellent.
\section{Conclusions and outlooks} \label{s:conc}
We have suggested a scheme to generate fully inseparable
three-mode entangled states of radiation based on interlinked
bilinear interactions taking place in a single $\chi^{(2)}$
nonlinear crystal. We have shown how the resulting three-mode
entanglement can be used to realize symmetric and antisymmetric
telecloning machines that achieve optimal fidelity for coherent
states. An experimental implementation involving a BBO nonlinear
crystal is suggested and the feasibility of the scheme is
analyzed. Preliminary experimental results are presented: as a
first quantitative check, we measured the energy of the beam
generated at $\omega_2$ as a function of the energy of the
ordinarily polarized pump, for fixed values of the energies of the
extraordinarily polarized pump-beam, and of the seed-beam. The
agreement between measured and calculated values is excellent.
\par\noindent
To realize the telecloning protocol described in Section
\ref{s:3cl} we need to generate three-mode entanglement from
vacuum. This should be possible by implementing the same
experimental setup as in Fig.~\ref{f:setup} with a different laser
source able to deliver a higher intensity. In fact, we plan to use
a mode-locked amplified Nd:YLF laser (IC-500, HIGH Q Laser
Production, Hohenems, Austria) with which it is easy to achieve an
intensity value of $50\ \mathrm{GW}/\mathrm{cm}^2$ in a collimated
beam. Since such a value was enough to generate bright twin beams
in a 4-mm thick BBO crystal, it should allow us to obtain the
three-mode entangled state described in this paper, not only by
seeding the crystal but also for initial vacuum state.
\section*{Acknowledgment}
This work has been supported by the INFM through the project
PRA-2002-CLON and by MIUR through the project FIRB-RBAU014CLC. 
The authors thank M. Cola and N. Piovella
(Universit\`a di Milano) for fruitful discussions, and F. Paleari
and F. Ferri (Universit\`a dell'Insubria) for experimental
support.

\appendix
\section{Heisenberg evolution of modes}
\label{a:hei}
In this section we calculate the dynamics generated
by the Hamiltonian (\ref{intH}) in the Heisenberg picture. The
equations of motion are given by
\begin{eqnarray}
  \label{EqMotion}
  \dot{a}_1^\dag &=& i\overline{\gamma}_1a_3 \nonumber \\
  \dot{a}_2 &=& -i\gamma_2a_3 \nonumber \\
  \dot{a}_3 &=& -i\gamma_1a_1^\dag-i\overline{\gamma}_2a_2 \:.
\end{eqnarray}
This system od differential equations can be Laplace transformed
in the following algebraic system
\begin{eqnarray} \label{EqMotionLap}
a_1^\dag(0)+\mu \tilde{a}_1^\dag (\mu) &=& i\overline{\gamma}_1\tilde{a}_3(\mu)
\nonumber \\ a_2(0)+\mu \tilde{a}_2(\mu) &=&
-i\gamma_2\tilde{a}_3(\mu) \nonumber \\ a_3(0)+\mu
\tilde{a}_3(\mu) &=&
-i\gamma_1\tilde{a}_1^\dag(\mu)-i\overline{\gamma}_2\tilde{a}_2(\mu) \:,
\end{eqnarray}
where we have defined the Laplace transform of $a_j(t)$
\begin{equation}
  \label{Lap}
  \tilde{a}_j(\mu)\equiv \int_0^\infty dt\: e^{-\mu t}a_j(t) \:.
\end{equation}
The determinant of the system (\ref{EqMotionLap}) is
\begin{equation}
  \label{det}
  \Delta=\mu(\mu+\Gamma)(\mu-\Gamma) \:,
\end{equation}
where $\Gamma\equiv\sqrt{|\gamma_1|^2-|\gamma_2|^2}$, therefore its
solution reads
\begin{eqnarray} \label{SolLap} \tilde{a}_1^\dag (\mu) &=&
\frac{1}{\Delta}\left[(|\gamma_2|^2+\mu^2)a_1^\dag
(0)+\overline{\gamma}_1\overline{\gamma}_2a_2(0)+i\overline{\gamma}_1\mu
a_3(0)\right] \nonumber \\ \tilde{a}_2(\mu) &=&
\frac{1}{\Delta}\left[-\gamma_1\gamma_2a_1^\dag
(0)+(\mu^2-|\gamma_1|^2)a_2(0)-i\gamma_2\mu
a_3(0)\right] \nonumber \\ \tilde{a}_3(\mu) &=&
\frac{1}{\Delta}\left[-i\gamma_1\mu a_1^\dag
(0)-i\mu\overline{\gamma}_2a_2(0)+\mu^2 a_3(0)\right] \:.  \end{eqnarray}
The solution of system (\ref{EqMotion}) follows from anti-transforming Eq.
(\ref{SolLap}). We have
\begin{eqnarray}
a_1^\dag (t) &=& f_1 a_1^\dag (0) + f_2 a_2  (0) + f_3 a_3 (0) \\
a_2 (t) &=& g_1 a_1^\dag (0) + g_2 a_2 (0) + g_3 a_3 (0) \\
a_3 (t) &=& h_1 a_1^\dag (0) + h_2 a_2 (0) + h_3 a_3 (0)
\end{eqnarray}
where the coefficients are given by
\begin{eqnarray}
f_1(t) &=& \frac{1}{\Omega^{2}}\left[|\gamma_1|^2 \cos{\Omega t} -|\gamma_2|^2\right]
\\ f_2(t) &=& \frac{\overline{\gamma}_1 \overline{\gamma_2}}{\Omega^{2}} \left[\cos{\Omega t}-1\right]
\\ f_3(t) &=& i\frac{\overline{\gamma}_1}{\Omega} \sin{(\Omega t)}
\\g_1(t) &=&\frac{\gamma_1 \gamma_2}{\Omega^{2}} \left[1-\cos{\Omega t}\right]
\\g_2(t) &=&\frac{1}{\Omega^{2}} \left[|\gamma_1|^2-|\gamma_2|^2 \cos{\Omega t}\right]
\\g_3(t) &=&-i\frac{\gamma_2}{\Omega} \sin{(\Omega t)}
\\h_1(t) &=&-i\frac{\gamma_1}{\Omega} \sin{(\Omega t)}
\\h_2(t) &=&-i\frac{\overline{\gamma_2}}{\Omega} \sin{(\Omega t)}
\\h_3(t) &=&\cos{(\Omega t)}
\label{fgh}\;
\end{eqnarray}
and $\Omega \equiv i\Gamma = \sqrt{|\gamma_2|^2 -|\gamma_1|^2}$.
\section{Schrodinger evolution in a seeded crystal}
\label{a:SchAlpha}
In this appendix we derive the explicit expression of the evolved state 
from $|\alpha,0,0\rangle$. We can write the 
Hamiltonian (\ref{intH}) as follows:
\begin{eqnarray}
  \label{HamKJ}
  H_{int}=\gamma_1K^\dag + \overline{\gamma_2}J + h.c. \:,
\end{eqnarray}
with the definitions $K\equiv a_1a_3$ and $J\equiv a_2a^\dag_3$.
To calculate the evolved state we can proceed by factorizing the
temporal evolution operator of the system; to this purpose we
introduce the following operators
$$
J_1\equiv a_1a^\dag_1+a^\dag_3a_3 \:, \qquad J_2\equiv
a^\dag_3a_3-a^\dag_2a_2 \:, \qquad M\equiv a_1a_2 \:, $$
which form with
K and J a closed algebra. Actually, the temporal evolution operator
can be written in the following way:
\begin{equation} \label{UdiTFact}
\hat{U}(t)=e^{\beta_1K^\dag}e^{\beta_2M^\dag}e^{\beta_3J^\dag}e^{\beta_4J_1}e^{\beta_5J_2}
e^{\beta_6J}e^{\beta_7K}e^{\beta_8M}
\:, \end{equation}
which allows us to calculate the evolution of a generic initial state
as a function of $\beta_i$. In the case under investigation we obtain:
\begin{widetext}
\begin{eqnarray} \label{EvStateBetas} \hat{U}(t)|\alpha,0,0\rangle &=&
\hat{U}(t)e^{-\frac{|\alpha|^2}{2}}\sum_{n}\frac{\alpha^n}{\sqrt{n!}}|n,0,0\rangle
\nonumber \\ &=&
e^{-\frac{|\alpha|^2}{2}}e^{\beta_1K^\dag}e^{\beta_2M^\dag}e^{\beta_3J^\dag}e^{\beta_4J_1}\sum_{n}
\frac{\alpha^n}{\sqrt{n!}}|n,0,0\rangle \nonumber \\ &=&
e^{-\frac{|\alpha|^2}{2}}e^{\beta_1K^\dag}e^{\beta_2M^\dag}e^{\beta_3J^\dag}e^{\beta_4}\sum_{n}
\frac{(\alpha e^{\beta_4})^n}{\sqrt{n!}}|n,0,0\rangle \nonumber \\ &=&
e^{-\frac{|\alpha|^2}{2}}e^{\beta_4}e^{\beta_1K^\dag}\sum_{n,p}\frac{(\alpha
e^{\beta_4})^n}{\sqrt{n!}}\frac{\beta_2^p}{\sqrt{p!}}\frac{\sqrt{(n+p)!}}{\sqrt{n!}}|n+p,p,0\rangle
\nonumber \\ &=&
e^{-\frac{|\alpha|^2}{2}}e^{\beta_4}\sum_{n,p,q}\beta_1^q\beta_2^p(\alpha
e^{\beta_4})^n\frac{\sqrt{(n+p+q)!}}{n!\sqrt{p!q!}}|n+p+q,p,q\rangle \:.
\end{eqnarray}
\end{widetext}
It can be demonstrated \cite{nic} that
$$
e^{\beta_4}=\frac{1}{\sqrt{1+N_1}}\:,\: \qquad
\beta_1=\sqrt{\frac{N_3}{1+N_1}}\:,\: \qquad
\beta_2=\sqrt{\frac{N_2}{1+N_1}} \:.$$
Moreover, for the population with initial vacuum
$N_j = \langle {\bf T}_0 | a^\dag_j a_j |{\bf T}_0\rangle$
and initial seed
$N_{j\alpha} = \langle {\bf T}_\alpha | a^\dag_j a_j |{\bf T}_\alpha \rangle$
we have the relations
$$ N_1=\frac{N_{1\alpha}-|\alpha|^2}{1+|\alpha|^2}\:,\: \qquad
N_2=\frac{N_{2\alpha}}{1+|\alpha|^2}\:,\: \qquad
N_3=\frac{N_{3\alpha}}{1+|\alpha|^2} \:.$$
\end{document}